\documentclass[4apaper,12pt]{article}

\usepackage[english]{babel}
\usepackage{amssymb}
\usepackage{graphicx,amsmath,amsfonts,amssymb,cite}

\newcommand{\s}{$s$\ }
\newcommand{\p}{$p_{1/2}$}
\newcommand{\cm}{HFI$_\mu$}
\newcommand{\cq}{HFI$_Q$}
\newcommand{\km}{QM$_\mu$}
\newcommand{\kq}{QM$_Q$}
\newcommand{\m}{m_e}
\newcommand{\E}{E_{n,1}}
\newcommand{\Em}{E_{n,-1}}

\newcommand{\e}{E_{n,1}}

\newcommand{\az}{\alpha Z}
\newcommand{\me}{m_e}

\newcommand{\ka}{\kappa}

\newcommand{\la}{\langle}
\newcommand{\ra}{\rangle}
\newcommand{\sx}{\sigma_x}

\newcommand{\sz}{\sigma_z}

\begin{document}
\title{ELECTRIC QUADRUPOLE MOMENT OF A HYDROGENLIKE ION IN $s$ AND $p_{1/2}$ STATES}
\date{}
\author{ Y.~S.~Kozhedub and
 V.~M.~Shabaev}
 \maketitle
 \begin{center}
Department of Physics, St.Petersburg State University,
Oulianovskaya 1, Petrodvorets, St.Petersburg 198504, Russia
 \end{center}
 \abstract{ Relativistic formulas for the electric quadrupole moment
 of a hydrogenlike atom, induced by the hyperfine interaction, are
 derived for $n$\s and $n$\p\ states. Both the magnetic dipole and
 electric quadrupole hyperfine interactions are taken into account.
 The formulas are valid for ions with arbitrary nuclear charge and
 spin. The induced quadrupole moment is compared with the nuclear
 quadrupole moment for a wide range of hydrogenlike ions.}
\section{Introduction}

    In Ref. \cite{Bar0} it was first noted that a hydrogen atom
    in the triplet \s state has an electric quadrupole moment. This is caused by the admixture of the
    $d_{3/2}$ states to the \s state due to the hyperfine interaction between
    the electron and the magnetic dipole moment of the nucleus (\cm).
    In Refs.\cite{Bar1,Bar2,Khrip1}, the quadrupole moment (\km) of a hydrogenlike atom in the $1$\s state
     was calculated in the full relativistic theory:
     \begin{equation}\label{1}
         Q_\mu(1s,Z,I,\mu)=\frac{2\gamma+1}{3} \frac{\eta}{Z} \frac{\mu}{\mu_{N}} Q_{1},
     \end{equation}
    Here
     \begin{equation}
      Q_{1}=\frac{1}{3} \alpha^{2} \frac{\m}{m_{p}} a^{2},
     \end{equation}
    $Z$ is the nuclear charge number, $I$ is the nuclear spin, $\alpha$ is the fine structure constant, $a$ is the Bohr radius, $\m$ and $m_{p}$ are the
    electron and proton masses, respectively, $\mu$ is the magnetic
    moment of the nucleus,
    $\mu_{N}=\vert e \vert\hbar/(2 m_{p}c)$ is the nuclear magneton,
    $\gamma=\lbrack1-(\az)^2\rbrack^{1/2}$,
        \begin{equation}
      \eta=(-1)^{F-I-1/2}\frac{F(2F-1)(2F+1)}{I(I+F+1/2)(I+F+3/2)}
     \end{equation}
    and $F$ is the total atomic angular momentum.
    For the $n$\s state with $F=1$, $I=1/2$ the corresponding
    calculation to the lowest order in $\az$ yields \cite{Khrip2}:
     \begin{equation}\label{5}
      Q_\mu(ns,1,1/2,\mu)=\frac{n^2+2}{3}\frac{\mu}{\mu_{N}}Q_{1}.
     \end{equation}

    A hydrogenlike atom has the \km\ also in \p\ states, which is caused by the admixture of  $p_{3/2}$ states. The relativistic
    theory of the \km\ for a hydrogenlike atom in the 2\p\ state with $F=1$, $I=1/2$ gives \cite{Bar3}:
     \begin{equation}\label{4}
       Q_\mu(2p_{1/2},Z,1/2,\mu)=-\frac{1}{15}\frac{N-1}{2-N}(4N^3+N^2+9N-6)\frac{1}{Z}\frac{\mu}{\mu_{N}}Q_{1},
     \end{equation}
     where \ \ \ \begin{math} N=(2+2\gamma)^{1/2}.
     \end{math}

     The electric quadrupole hyperfine interaction of the electron
     with the nucleus (\cq)\ can also induce a nonzero electric quadrupole
     moment of the atom (\kq), mixing the \s states with the $d$ states,
     and the \p\ states with the $p_{3/2}$ and  the $f_{5/2}$ states. However,
     since for light atoms the \cq\ is much weaker than the \cm\
     and, in addition, it can contribute only for atoms with the nuclear spin
     $I>1/2$, it was generally disregarded in previous calculations.

     In the present paper, we derive relativistic
 expressions for the
electric quadrupole moment of a hydrogenlike ion being
 in  $ns$ and $np_{1/2}$ states. The
derived formulas are valid for ions
 with arbitrary nuclear charge, spin, magnetic dipole and
    electric quadrupole moments.
 Both  magnetic dipole and  electric quadrupole hyperfine
interactions are
    taken into account.

     Relativistic and Heaviside charge units
($\hbar=c=1~,\; \alpha=e^{2}/4\pi$) are used in the paper, the
charge of the electron is taken to be $e<0$.
\section{Induced quadrupole moment}

    The hyperfine interaction operator is given by the sum
\begin{equation}
H_{\rm HFI}=H_\mu+H_Q,
\end{equation}
where $H_\mu$ and $H_Q$ are the magnetic-dipole and
electric-quadrupole hyperfine interaction operators, respectively.
In the point-dipole approximation,
\begin{equation}\label{FB}
H_\mu=\frac{|e|}{4\pi}\frac{(\vec{\alpha}\cdot[\vec{\mu}\times
\vec{r}])}{r^3}\,,
\end{equation}
and, in the point-quadrupole approximation,
\begin{equation}
H_Q = -\alpha \sum_{m=-2}^{m=2}Q_{2m} \eta_{2m}^*(\vec{n})\,.
\end{equation}
Here the vector $\vec{\alpha}$ incorporates the Dirac $\alpha$
matrices, $\vec{\mu}$ is the nuclear magnetic moment operator
acting in the space of nuclear variables, $
Q_{2m}=\sum_{i=1}^Zr_i^2C_{2m}(\vec{n}_i)$ is the operator of the
electric quadrupole moment of the nucleus,
$\eta_{2m}=C_{2m}(\vec{n})/r^3 $ is an operator that acts on
electron variables, $ \vec{n}=\vec{r}/r$, $\vec{r}$ is the
position vector of the electron, $\vec{r}_i$ is the position
vector of the $i$th proton in the nucleus, $C_{lm}=
\sqrt{4\pi/(2l+1)}\,Y_{lm}$, and $Y_{lm}$ is a spherical harmonic.
It must be stressed that the electric quadrupole interaction
should be taken into account only for ions with $I > 1/2$.
 Thus the quadrupole moment of a
 hydrogenlike atom in the $|A\ra$ state, induced by the hyperfine interaction, is given by
\begin{equation}\label{Q1}
    Q_{\mu}(A)+Q_Q(A)=2\sum^{ E_{N} \neq E_{A} }_{N}
        \frac {\la A|Q_{zz}|N\ra\la N|H_{\mu}+H_Q|A\ra}
        {E_{A}-E_{N}},
        \qquad (M_{ F_{ A } }=F_{A}),
\end{equation}
where $Q_{zz}=-r^2(3n_{z}^2-1)$, $|A\ra$ and $|N\ra$ are the state
vectors of the total (electron plus nucleus) atomic system, $E_A
~\mbox{and}\ E_N$ are the related energies.

Let us consider first the $n$\s state. Integrating over angular
variables in (\ref{Q1}) yields
     \begin{equation}\label{Q2}
     Q_{\mu}(ns,Z,I,\mu)=\frac{2e}{15\pi} \eta\mu
     \la n-1|r^2|\xi_1(2;n,-1)\ra,
    \end{equation}
     and
     \begin{equation}\label{qq3}
    Q_Q(ns,Z,I,Q_N)=\alpha\tau Q_N(2\la n-1|r^2|\xi_2(2;n,-1)\ra+3\la n-1|r^2|\xi_2(-3;n,-1)\ra).
    \end{equation}
    Here $\mu=\la II|\mu_z|II\ra$ is the nuclear magnetic moment,
    $Q_N=2\la II|Q_{20}|II\ra$ is the electric quadrupole moment of
    the nucleus,
    \begin{equation}\label{tay}
     \tau= \left\{
     \begin{array}{cl}
     \frac{2}{25} & ,\ F=I+1/2, I\neq1/2\\
     \frac{2(I-1)(2I+3)}{25I(2I+1)} & ,\ F=I-1/2, I\neq1/2 \\
     0 & ,\ I=1/2
     \end{array} \right .,
    \end{equation}

\begin{equation}\label{sum1}
    |\xi_1(\ka';n,\ka)\ra \equiv \sum_{n^{\prime}}^{(E_{n',\ka'} \neq E_{n,\ka})}
    \frac{|n'\ka'\ra\la
    n'\ka'|\sigma_xr^{-2}|n\ka\ra}{E_{n,\ka}-E_{n^{\prime},\ka^{\prime}}},
    \end{equation}
     \begin{equation}\label{sum2}
    |\xi_2(\ka';n,\ka)\ra \equiv \sum_{n^{\prime}}^{(E_{n',\ka'} \neq E_{n,\ka})}
    \frac{|n'\ka'\ra\la
    n'\ka'|r^{-3}|n\ka\ra}{E_{n,\ka}-E_{n^,{\prime}\ka^{\prime}}},
    \end{equation}
     $\sigma_x$ is the Pauli matrix, the vector $|n\ka\ra =\Big( \begin{array}{c} rg_{n\ka}\\ rf_{n\ka} \end{array}
        \Big)$ $(\la n\ka|n\ka\ra=\int_0^{\infty}(g_{n\ka}^2+f_{n\ka}^2)r^2\, dr=1)$
consists of the upper and lower radial components of the Dirac
wave function defined by
    \begin{equation} \label{dir1}
                |n\ka m\ra=\Bigg( \begin{array}{c}
                g_{n\ka}(r)\Omega_{\ka m}(\vec{n})\\
                if_{n\ka}(r)\Omega_{-\ka m}(\vec{n})\\
                    \end{array} \Bigg),
     \end{equation}
    $\ka=(-1)^{j+l+1/2}(j+1/2)$ and $E_{n,\ka}$ is the Dirac
    energy.
    For the point-charge nucleus, the sums $\xi_1$ and $\xi_2$,
 can be evaluated analytically, employing the
method of  generalized virial relations for the Dirac equation in
a central field \cite{sha91,sha03}. For $\ka\neq\pm\ka'$ one can
derive \cite{sha91,sha03}
      \begin{align}\label{sum10}
|\xi_1(\ka',n\ka)\ra=&\{ [1-(\ka-\ka')^2][1-(\ka+\ka')^2]+4(\az)^2\}^{-1}\notag\\
&\times \biggl[[1-(\ka+\ka')^2]\biggl(\frac{4\az\m}{\ka^2-\ka'^2}+
(\ka'-\ka)r^{-1}+r^{-1}\sigma_{z}\notag\\
&-\frac{2}{\ka+\ka'}(\m\sigma_{x}+E_{n,\ka}i\sigma_{y})\biggr)+\frac
{4\az[(\ka+\ka')\m-E_{n,\ka}]}{\ka-\ka'}\notag\\
&+2\az[\sigma_{x} r^{-1}+(\ka+\ka')r^{-1}i\sigma_{y}]\biggr]
|nk\ra.
\end{align}
and \cite{Mosk}
    \begin{align}\label{sum20}
    |\xi_2(\ka',n\ka) \rangle
 =& \{[4-(\kappa-
\kappa')^2][4-(\kappa+\kappa')^2]+16(\az)^2\}^{-1} \biggl[ 2\az
(\kappa^2-\kappa'^2)\frac{1}{r^2}\notag  \\
&+(\kappa+\kappa')[4- (\kappa-\kappa')^2] \frac{\sigma_x}{r^2} +
(8+8(\az)^2
-2(\kappa-\kappa')^2) \frac{i\sigma_y}{r^2}\notag  \\
&-4\az(\kappa+\kappa')\frac{\sigma_z}{r^2} +\frac
{(\kappa'-\kappa)D_1}{D_0}\frac{1}{r}+\frac{D_2}{D_0}
\frac{\sigma_x}{r} +\biggl(\frac{D_2(\kappa+\kappa')}{D_0} \notag\\
&-8\az \me(\kappa+\kappa')\biggr)\frac{i\sigma_y}{r}
+\frac{D_1}{D_0} \frac{\sigma_z}{r}
-\frac{2}{\kappa+\kappa'}\frac{D_1}{D_0} (\me\sigma_x +
E_{n, \kappa}i\sigma_y)\notag   \\
&+\frac{4\az \me} {\kappa^2-{\kappa'}^2}\frac{D_1}{D_0}+
\frac{2\me(\kappa+\kappa')-2E_{n,\kappa}}
{\kappa-\kappa'}\frac{D_2}{D_0}  -\frac{16\az
\me^2(\kappa+\kappa')}{\kappa-\kappa'} \biggr]|n \kappa \rangle ,
    \end{align}
 where  $\sigma_y$ and $\sigma_z$ are the Pauli matrices,
   \begin{align}
D_0= &[1-(\kappa+\kappa')^2][1-(\kappa-\kappa')^2] +4(\az)^2,
\\   D_1=&2\me[8(\az)^2(\kappa+\kappa')^2 +
(1-(\kappa+\kappa')^2)(8+8(\az)^2-2(\kappa-\kappa')^2)] \notag\\ &
+2E_{n\kappa}(\kappa+\kappa')
[8(\az)^2+(1-(\kappa+\kappa')^2)((\kappa-\kappa')^2-4)],
\\
D_2=&8\az \me[4+4(\az)^2-(\kappa-\kappa')^2 -
(\kappa+\kappa')^2(1-(\kappa-\kappa')^2)] \notag
\\ &+12\az E_{n,\kappa}(\kappa+\kappa')
[(\kappa-\kappa')^2-2].\label{Do}
\end{align}
Further calculations of expressions (\ref{Q2}), (\ref{qq3}) can
easily be performed by using the recurrent formulas for the
expectation values $A^s=\la n\ka| r^{s}|n\ka\ra$, $B^s=\la
n\ka|\sz r^{s}|n\ka\ra$, and $C^s=\la n\ka|\sx r^{s}|n\ka\ra$
\cite{eps62,sha91,sha03}.

    Finally, for the $n$\s state, we obtain
    \begin{align}\label{Q3}
       Q_\mu(ns,Z,I,\mu)=& \eta\frac{4\me}{5Z} \frac{\mu}{\mu_N} \Big\{ \frac{1}{ 6( \m+\Em ) }
          \Big( \frac{ 4\Em^2+\m^2 }{ \m+\Em } \frac{ (\az)^2 }{
          \m-\Em}\notag\\
        &+\frac{3\Em\m-2\Em^2}{\m^2} \Big)+\frac{2\Em+\m}{4\m^2}
     \Big\}Q_1
    \end{align}
    and
    \begin{align}\label{qqf1}
    Q_Q(ns,Z&,I,Q_N)=\frac{5\tau Q_N}{12Z\me^2(15-16(\az)^2)(45+4(\az)^2)(\me^2-\Em^2)}\notag\\
     &\times \Big\{
     10(-1728\Em^4-3537\me\Em^3+6237\Em^2\me^2+1809\Em\me^3\notag\\
     &-2457\me^4)+\frac{3(\az)^2}{\me^2-\Em^2}
     \big(7405\me^6-736\Em^6+10528\me\Em^5\notag\\
     &-31033\Em^2\me^4+2406\Em^3\me^3-3934\Em\me^5+20764\Em^4\me^2\big)\notag\\
     &+\frac{(\az)^4\me}{\me^2-\Em^2} \big(
     6505\me^5-1120\Em^3\me^2+2240\Em\me^4+1280\Em^5\notag\\
     &+7872\me\Em^4-1612\Em^2\me^3\big)+
     \frac{400(\az)^6\me^4}{\me^2-\Em^2} \big( \me^2+4\Em^2 \big)
     \Big\}.
    \end{align}
    It can be seen that formula (\ref{1}) is a
    particular case ($n=1$, $F=1$, $I=1/2$ $(\eta=1)$) of  formula (\ref{Q3}).

For small $Z$, we can expand (\ref{Q3}) and (\ref{qqf1}) in the
parameter $\az$ with the two lowest-order terms kept:
     \begin{equation}\label{Q4}
    Q_\mu(ns,Z,I,\mu)=\frac{\eta}{3Z}  \frac{\mu}{\mu_{N}}
     \Big\{ (n^2+2)-(\alpha Z)^2
    \big(n-\frac{9}{20}\big(1-\frac{1}{n^2}\big)\big)+O((\az)^4)\Big\}Q_{1}.
    \end{equation}
    \begin{align}\label{tqq1}
    Q_Q(ns&,Z,I,Q_N)=\frac{\tau Q_N}{Z}\Big\{ \big( \frac{61}{36}n^4-\frac{355}{36}n^2+4
    \big) \notag\\
    &+\frac{(\az)^2}{540n^2}\big(1028n^6-1830n^5-4322n^4+5325n^3+3957n^2-2808
    \big)+O((\az)^4) \Big\}.
    \end{align}
    The main term in equation (\ref{Q4}) coincides with formula (\ref{5}).

     For the $n$\p\ state, a similar calculation yields
\begin{align}\label{Q6}
        Q_\mu(np_{1/2},Z,I,\mu)=&-\eta\frac{4\me}{5Z}\frac{\mu}{\mu_N}
       \Big\{ \frac{1}{6}\Big( \frac{4\E^{2}+\m^2}{\m+\E}
       \frac{(\az)^2}{(\m-\E)^2}\notag\\
       &-\frac{ 2\E^{2}+3\m\E }{\m^2} \frac{1}{\m-\E} \Big)-\frac{2\E-\m}{4\m^2} \Big\}Q_1.
\end{align}
\begin{align}\label{qqf2}
    Q_Q(np_{1/2}&,Z,I,Q_N)=\frac{5\tau Q_N}{12Z\me^2(15-16(\az)^2)(45+4(\az)^2)(\me-\E)}\notag\\
     &\times \Big\{ 54\big(-76\e^3+311\me\e^2+6\me^2\e-141\me^3\big)
     \notag\\
     &+\frac{3(\az)^2}{(\me^2-\e^2)(\me+\e)}\big(11805\me^6-41297\e^2\me^4\notag\\
     &-6112\me\e^5+10710\e\me^5+21132\e^4\me^2-998\e^3\me^3-640\e^6\big)\notag\\
     &+\frac{(\az)^4\me^2}{(\me^2-\e^2)(\me+\e)}\big(7785\m^4+1248\me\e^3-288\e\me^3\notag\\
     &-4460\e^2\me^2+8000\e^4\big)+\frac{400(\az)^6\me^4}{(\me^2-\e^2)(\me+\e)}\big(\me^2+4\e^2\big)
     \Big\}.
    \end{align}
    For $n=2$, $F=1$, $I=1/2$  formula (\ref{Q6}) agrees with equation (\ref{4}).
    For small $Z$, we can expand expressions (\ref{Q6}) and (\ref{qqf2}) in the
parameter $\az$ with the two lowest-order terms kept:
    \begin{align}\label{Q7}
    Q_\mu(np_{1/2},Z,I,\mu)=&-\frac{\eta}{3Z}  \frac{\mu}{\mu_{N}}
      \frac{1}{(\az)^2} \Big\{ 4(n^4-n^2)\notag\\
        &-(\alpha Z)^2(8n^3-\frac{19}{5}n^2-4n+\frac{4}{5}) +O((\az)^4)  \Big\} Q_{1}.
    \end{align}
    \begin{align}\label{tqq2}
    Q_Q(np_{1/2},Z,I,Q_N)=&-\frac{\tau Q_N}{(\az)^2Z}\Big\{\frac{20}{3}n^2(n^2-1)\notag\\
    &+\frac{(\az)^2}{12} \big(65n^4-160n^3-51n^2+80n+100
    \big)+O((\az)^4) \Big\}.
    \end{align}

\section{Numerical results}

    In Tables $1$ and $2$, we present the numerical results for
    $Q_{\mu}$, $Q_Q$, and $Q_{\textrm{total}}=\frac{25\tau}{2}Q_N+Q_\mu+Q_Q$ (the total
    quadrupole moment of the atom) for a wide range of hydrogenlike
    ions. For ions with $I=1/2$, the total quadrupole moment is
    completely determined by \km. For the other ions, the role of
    the induced quadrupole moment is most important for low $Z$
    and decreases with $Z$ increasing.

    We note also that, according to formulas (\ref{Q4}), (\ref{Q7}), (\ref{tqq1}), (\ref{tqq2}), the
    induced quadrupole moment increases rapidly with $n$
    increasing. As a result, for highly exited states the total
    quadrupole moment of a hydrogenlike atom is mainly determined by
    the induced quadrupole moment.

\clearpage
\begin{table}
\caption{The numerical results for the \km, the \kq, and
$Q_{\textrm{total}}$ in case of the $1s$ state. The values of
$\mu/\mu_N$ and $Q_N$ are taken from \cite{rag89}.}
    \begin{center}
    \begin{tabular}{|c|c|c|c|c|}
          \hline
      Ion                 & $^1\textrm{H}$ & $^{13}\textrm{C}^{5+}$ & $^{17}\textrm{O}^{7+}$ & $^{43}\textrm{Ca}^{19+}$  \\
      \hline
      Z                   & 1 & 6 & 8 & 20  \\
      \hline
      I                   & 1/2   & 1/2 & 5/2 & 7/2 \\
    \hline
    $\mu/\mu_N$   & 2.79285 & 0.702412(2) & -1.8938(1) & -1.3176  \\
    \hline
      $Q_N$, barn               & 0     & 0     & -0.02578 & -0.049(5)  \\
    \hline\hline
      $Q_\mu,F=I+1/2$, barn    & 0.7560 & 0.03167 & -0.06401 & -0.01771  \\
    \hline
      $Q_Q,F=I+1/2$, barn       & 0     & 0     & 0.001072 & 0.0008(1)  \\
    \hline
      $Q_{\textrm{total}},F=I+1/2$, barn & 0.7560 & 0.03167 & -0.08872 & -0.066(5)  \\
    \hline \hline
      $Q_\mu,F=I-1/2$, barn     & 0     & 0     & 0.02560 &0.009485 \\
    \hline
      $Q_Q,F=I-1/2$, barn       & 0     & 0     & 0.0008576 & 0.0007(1)  \\
    \hline
      $Q_{\textrm{total}},F=I-1/2$, barn & 0     & 0     & 0.00583 & -0.033(5) \\
      \hline
    \end{tabular}
\end{center}
    \begin{center}
    \begin{tabular}{|c|c|c|c|c|}
    \hline
      Ion                 & $^{131}\textrm{Xe}^{53+}$ &$^{207}\textrm{Pb}^{81+}$& $^{209}\textrm{Bi}^{82+}$ & $^{235}\textrm{U}^{91+}$ \\
      \hline
      Z                      & 54      &82  & 83            & 92 \\
      \hline
      I                    & 3/2   &1/2    & 9/2 & 7/2 \\
    \hline
    $\mu/\mu_N$  & 0.691862(4) &0.592583(9)& 4.1106(2) & -0.39(7)$^1$ \\
    \hline
      $Q_N$, barn              & -0.120(12)&0    & -0.50(8)     & 4.936(6) \\
    \hline\hline
      $Q_\mu,F=I+1/2$, barn     &  0.003281  &0.001697     & 0.01158       & -0.0009(2) \\
    \hline
      $Q_Q,F=I+1/2$, barn        & 0.0007(1)  &0       & 0.0014(2)       & -0.01095(1) \\
    \hline
      $Q_{\textrm{total}},F=I+1/2$, barn  & -0.12(1) &0.001697         & -0.49(8)       & 4.924(6) \\
    \hline \hline
      $Q_\mu,F=I-1/2$, barn     & -0.0005469  &0        & -0.007206      & 0.0005(1) \\
    \hline
      $Q_Q,F=I-1/2$, barn           & 0.00033(3) &0         & 0.0013(2) & -0.00978(1) \\
    \hline
      $Q_{\textrm{total}},F=I-1/2$, barn     & -0.06(1)&0      & -0.47(7) & 4.398(5)\\
      \hline
    \end{tabular}
    \end{center}
        $^1$ An average of the values given in \cite{rag89}.
\end{table}

\begin{table}
\caption{ The numerical results for the \km, the \kq, and
$Q_{\textrm{total}}$ in case of the $2$\p\ state. The values of
$\mu/\mu_N$ and $Q_N$ are taken from \cite{rag89}.}
\begin{center}
\begin{tabular}{|c|c|c|c|c|}
      \hline
      Ion                 & $^1\textrm{H}$ & $^{13}\textrm{C}^{5+}$ & $^{17}\textrm{O}^{7+}$ & $^{43}\textrm{Ca}^{19+}$  \\
      \hline
      Z                   & 1 & 6 & 8 & 20  \\
      \hline
      I                   & 1/2   & 1/2 & 5/2 & 7/2  \\
    \hline
    $\mu/\mu_N$   & 2.79285 & 0.702412(2) & -1.8938(1) & -1.3176  \\
    \hline
      $Q_N$, barn               & 0     & 0     & -0.02578 & -0.049(5)  \\
    \hline\hline
      $Q_\mu,F=I+1/2$, barn     & -227170(100) & -264.1 & 300.0& 13.15    \\
    \hline
      $Q_Q,F=I+1/2$, barn       & 0     & 0     &6.051   &0.73(8)      \\
    \hline
      $Q_{\textrm{total}},F=I+1/2$, barn & -227170(100)     & -264.1    & 306.0   &13.83(8)     \\
    \hline \hline
      $Q_\mu,F=I-1/2$, barn     & 0     & 0     & -120.0 & -7.044   \\
    \hline
      $Q_Q,F=I-1/2$, barn       & 0     & 0     & 4.838 &0.65(8)\\
    \hline
      $Q_{\textrm{total}},F=I-1/2$, barn & 0     & 0     & -115.2 &-6.44(8) \\
      \hline
    \end{tabular}
\end{center}
\begin{center}
    \begin{tabular}{|c|c|c|c|c|}
      \hline
      Ion                & $^{131}\textrm{Xe}^{53+}$ &$^{207}\textrm{Pb}^{81+}$& $^{209}\textrm{Bi}^{82+}$ & $^{235}\textrm{U}^{91+}$ \\
      \hline
      Z                           & 54       &82 & 83            & 92 \\
      \hline
      I                        & 3/2      &1/2 & 9/2           & 7/2 \\
    \hline
    $\mu/\mu_N$    & 0.691862(4) &0.592583(9)& 4.1106(2) & -0.39(7)$^1$ \\
    \hline
      $Q_N$, barn           & -0.120(12)&0 & -0.50(8)   & 4.936(6) \\
    \hline\hline
      $Q_\mu,F=I+1/2$, barn       & -0.3096&  -0.06061  & -0.4015    & 0.025(5) \\
    \hline
      $Q_Q,F=I+1/2$, barn        & 0.09(1) & 0 & 0.10(3)   & -0.723(1) \\
    \hline
      $Q_{\textrm{total}},F=I+1/2$, barn      &-0.34(1) &   -0.06061&-0.8(1)     &4.24(1) \\
    \hline \hline
      $Q_\mu,F=I-1/2$, barn       & 0.05160  & 0 & 0.2598   & -0.013(5)\\
    \hline
      $Q_Q,F=I-1/2$, barn        & 0.045(5) & 0 & 0.093(15)   & -0.645(1)\\
    \hline
      $Q_{\textrm{total}},F=I-1/2$, barn     & 0.04(1) &  0& -0.11(9)   & 3.75(1)\\
      \hline
    \end{tabular}
\end{center}
$^1$ An average of the values given in \cite{rag89}.
\end{table}

\end{document}